# Strain effects on monolayer MoSi$_2$N$_4$: ideal strength and failure mechanism

Qingfang Li[1*], Wanxin Zhou[1], Xiangang Wan[3,4], Jian Zhou[2*]

[1] *Department of Physics, Nanjing University of Information Science & Technology, Nanjing 210044, China*

[2] *National Laboratory of Solid State Microstructures and Department of Materials Science and Engineering, Nanjing University, Nanjing 210093, China*

[3] *National Laboratory of Solid State Microstructures and School of Physics, Nanjing University, Nanjing 210093, China*

[4] *Collaborative Innovation Center of Advanced Microstructures, Nanjing University, Nanjing 210093, China*

* Email addresses: qingfangli@nuist.edu.cn, zhoujian@nju.edu.cn

Abstract: Recently, two-dimensional monolayer MoSi$_2$N$_4$ with hexagonal structure was successfully synthesized in experiment (Hong *et al* 2020 *Science* **369**, 670). The fabricated monolayer MoSi$_2$N$_4$ is predicted to have excellent mechanical properties. Motived by the experiment, we perform first-principles calculations to investigate the mechanical properties of monolayer MoSi$_2$N$_4$, including its ideal tensile strengths, critical strains, and failure mechanisms. Our results demonstrate that monolayer MoSi$_2$N$_4$ can withstand stresses up to 51.6 and 49.2 GPa along zigzag and armchair directions, respectively. The corresponding critical strains are 26.5% and 17.5%, respectively. For biaxial strain, the ideal tensile strength is 50.6 GPa with a critical strain of 19.5%. Compared with monolayer MoS$_2$, monolayer MoSi$_2$N$_4$ possesses much higher elastic moduli and ideal tensile strengths for both uniaxial and biaxial strains. Interestingly, the critical strain and failure mechanism of zigzag direction in MoSi$_2$N$_4$ are almost the same as those of armchair direction in MoS$_2$, while the critical strain and failure mechanism of armchair direction for MoSi$_2$N$_4$ are similar to the ones of zigzag direction for MoS$_2$. Our work reveals the remarkable mechanical characteristics of monolayer MoSi$_2$N$_4$.

**I Introduction**

Since the first isolation of two-dimensional (2D) graphene from graphite[1], tremendous efforts have been devoted toward developing and using 2D materials[2-4]. Among these 2D materials, MXenes, belonging to a novel transition metal carbide/nitride family, have great application potential [5]. Up to now, more than 30 stoichiometric MXenes have been prepared[5-7]. MXenes possess intriguing physical and chemical properties such as the high surface area to volume ratio, great miscibility, surface charge state, accessible active sites, and electron-rich density[8]. Accompanied by the rapid development of MXenes, researchers have increasingly unearthed their exciting applications, including energy storage, electromagnetic interference shielding, gas sensors, catalysis, antennas, supercapacitors, and transparent conducting films[9-14].

Very recently, a novel hexagonal 2D MXene, monolayer $MoSi_2N_4$ with excellent ambient stability, has been successfully synthesized by chemical vapor deposition[15]. Monolayer $MoSi_2N_4$ is an indirect semiconductor with a moderate band gap (1.94 eV)[15]. Besides, it is predicted that the electron and hole mobilities of monolayer $MoSi_2N_4$[15] are 4 to 6 times larger than those of monolayer $MoS_2$[16]. The moderate band gap, high mobility, and excellent stability of monolayer $MoSi_2N_4$ promise its advantages in nano-electronic and optoelectronics. Since the experimental synthesis of layered $MoSi_2N_4$, an enormous amount of research efforts have been devoted to exploring its various properties[17-27]. Density functional theory (DFT) calculations indicated that monolayer $MoSi_2N_4$ has a larger Poisson's ratio (0.28)[17] than that of graphene (0.16)[28]. Guo *et al* investigated the effects of biaxial strain on electronic, transport, and piezoelectric properties of monolayer $MoSi_2N_4$[20]. Their results demonstrate that the increasing tensile strain can enhance the piezoelectric stress coefficient, and the compressive strain can change the numbers of conduction band extrema[20].

Because of its high Young's modulus (about 1000 GPa)[29], graphene is considered as the strongest material which has potential applications in protective coating, structural composites, fibers, etc[30-32]. However, many of its record-breaking properties were measured on small samples, and pristine graphene has poor solubility in the conventional solvents[33]. Therefore, it is desirable to find more 2D materials with promising elastic characteristics. Monolayer $MoSi_2N_4$ exhibits excellent mechanical properties. For example, the measurement of mechanical properties demonstrated that monolayer $MoSi_2N_4$ has high Young's modulus and breaking strength[15], and the corresponding values based on DFT calculations (~ 479 and 49 GPa)[15] are much higher than those of monolayer $MoS_2$ (270 and 22 GPa)[34] and the other MXenes such as monolayer $Ti_3C_2T_x$ (333 and 17 GPa)[33] and $Nb_4C_3T_x$ (386 and 26 GPa)[35].

The adventitious strain is unavoidable, but the practical strain should not exceed the critical strain. For 2D materials, the critical strain and ideal tensile stress are crucial mechanical parameters which characterize the elastic limit of thin films and the nature of their chemical bonds[36,37]. The elastic limit and related mechanical properties of well-known 2D materials have been widely concerned, such as graphene[29,38-40], monolayer $MoS_2$[34,41-43], borophene[44,45], black phosphorene[46,47], h-BN[48-50], silicene[51], and $Ti_3C_2O_2$[52]. Therefore, it is very important to understand the elastic limit and the underlying mechanism of monolayer $MoSi_2N_4$. In this paper, we performed systematic analyses on the tensile strain-induced mechanical properties of monolayer $MoSi_2N_4$, including the critical strain, the ultimate stress, the change of buckling heights and bond lengths, and its failure mechanism as approaching the limit tensile strain.

## II Computational methods

All the structural optimization is performed by using the Vienna ab initio simulation package(VASP)[53,54] based on DFT. The generalized gradient approximation (GGA) with the Perdew-Burke-Ernzerh of exchange-correlation functional[55] and projected augmented wave (PAW) method[56,57] are used. A plane-wave cutoff energy of 520 eV and k-mesh of 15×15×1 are used. Both the internal atomic coordinates and lattice constants are relaxed until the Hellmann-Feynman forces become less than $10^{-4}$ eV/Å. All phonon dispersions are obtained by using the PHONOPY code[58]. The 4×4×1 and 4×2×1 supercells for the hexagonal primitive cell and orthogonal supercell in Fig. 1 are used to calculate the phonon frequencies.

## III Results and discussions

The optimized structure of monolayer $MoSi_2N_4$ is shown in Fig. 1. In the hexagonal primitive cell, there are four nitrogen, two silicon, and one molybdenum atom. Monolayer MoSi$_2$N$_4$ can be viewed as a 2H-MoS$_2$-like MoN$_2$ monolayer sandwiched in-between two slightly buckled honeycomb SiN layers. MoN$_2$ layer is bonded to SiN layers via vertical Si-N bonds with Mo atom located at the center of the trigonal building block of six N atoms. The optimized lattice constants are a=b=2.911 Å, which are in excellent agreement with the previous results[15,17]. The calculated buckling heights of MoN$_2$ and SiN layers are 2.506 and 0.506 Å, respectively. The d$_1$ (d$_2$) and d$_3$ (d$_4$) bond lengths are 2.096 and 1.755 Å, respectively. The bond angles $\theta_1(\theta_2)$ and $\theta_3(\theta_4)$ are 87.95 and 112.05º. These results are well consistent with the previous reports[17].

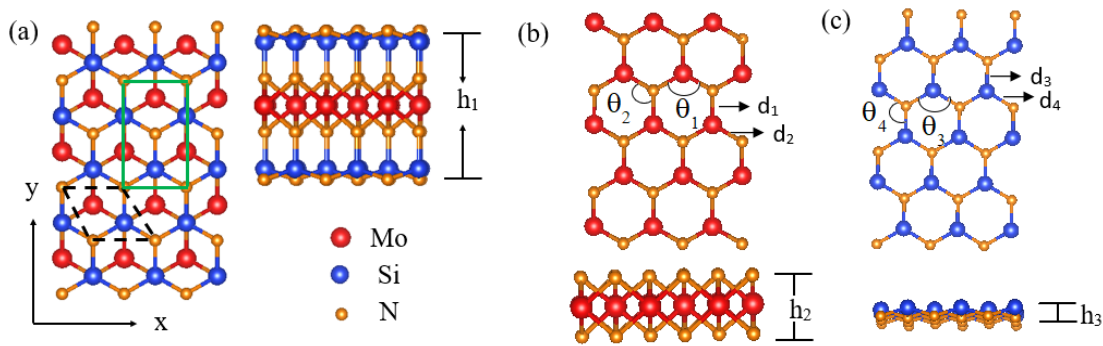

Fig.1 Top and side views of the crystal structures of (a) monolayer MoSi$_2$N$_4$, (b) MoN$_2$ layer, and (c) SiN layer. The black dashed lines represent the hexagonal primitive cell, and the green lines represent the orthogonal supercell.

The tensile strains of monolayer MoSi$_2$N$_4$ are applied in either biaxial or uniaxial (x and y) directions to check its ideal tensile strength. The calculated stress-strain relations of monolayer MoSi$_2$N$_4$ are illustrated in Fig. 2. By fitting the initial stress-strain curve under a small uniaxial strain of 1%, the corresponding elastic moduli are $E_x$=448.3±5.1 GPa and $E_y$=457.8±3.9 GPa, which are more than twice those of MoS$_2$ ($E_x$=197.9±4.3 GPa and $E_y$=200±3.7[42]). The degeneracy of elastic moduli means that the monolayer MoSi$_2$N$_4$ is an almost elastic isotropic 2D material. The elastic properties of monolayer MoSi$_2$N$_4$ have been measured by using atomic force microscope (AFM) as an indentor applied on the MoSi$_2$N$_4$ monolayer suspended on the SiO$_2$/Si substrate with the array of holes[15]. Under such experimental conditions, monolayer MoSi$_2$N$_4$ is most likely under biaxial tensile stress. The experiment obtained a Young's modulus of 491.4±139.1 GPa for MoSi$_2$N$_4$ monolayer[15]. Our calculations predict that the biaxial elastic modulus (599.5±7.2 GPa) is in good agreement with the experimental data and previous theoretical report (0.6 TPa[17]). The value is much larger than that of monolayer MoS$_2$ (250.2±5.8 GPa[42]), while smaller than that of graphene (1050 GPa[38]). For the in-plane uniaxial strain, the degenerate elastic responses of x (zigzag) and y (armchair) directions in a small range attribute to the symmetry of the hexagonal structure. The stress-strain relations become nonlinear with the increasing of uniaxial strains, and the stress under the uniaxial strain along armchair direction enhances faster than the one along zigzag direction. It can be argued the in-plane symmetry of monolayer MoSi$_2$N$_4$ is broken when the structure deviates dramatically from the hexagonal lattice under large uniaxial strains. In contrast, the stress-strain relation under the biaxial strain shows identity along the zigzag and armchair directions because the hexagonal symmetry is always kept.

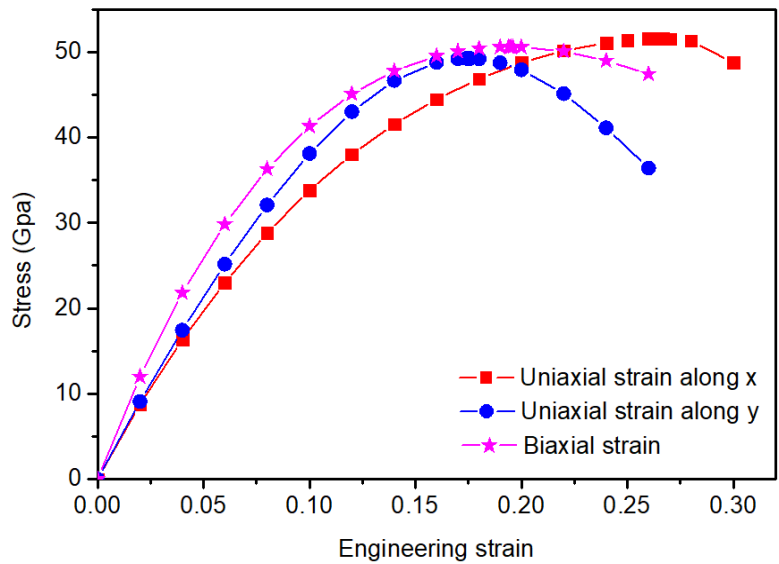

Fig.2 Stress-strain relationships of monolayer MoSi$_2$N$_4$ under uniaxial x, y, and biaxial tensions.

The calculated ideal strengths and critical strains of monolayer MoSi$_2$N$_4$ under three types of tensile strains are summarized in Table I. We also compare the corresponding values and failure mechanisms of MoSi$_2$N$_4$ with those of monolayer MoS$_2$ listed in Table I. The maximum stress for the uniaxial tension along the zigzag direction is 51.6 GPa, and the corresponding critical strain is 0.265. The monolayer MoSi$_2$N$_4$ can sustain more tensile strain along the zigzag direction than the armchair direction, and the ideal strength along the zigzag direction is slightly larger than that along the armchair direction. The ideal strength of 50.6 GPa under biaxial strain is closed to the ones under uniaxial strains, and the corresponding critical tensile strain is 0.195. The calculated ideal strength is in good agreement with the previous data[15,24]. Interestingly, the critical strain in $MoSi_2N_4$ under biaxial strain is the same as that of monolayer $MoS_2$ which is ascribed to their similar hexagonal structures. The critical strains of $MoSi_2N_4$ monolayer under zigzag and armchair tensile strains are 0.265 and 0.175 respectively, which are close to those of $MoS_2$ monolayer (0.256 and 0.18 along the armchair and zigzag directions). Obviously, their direction with similar critical stress is different. In addition, the tensile strengths of monolayer $MoSi_2N_4$ are more than twice those of $MoS_2$ for tree strain paths, which is mainly attributed to the strong interactions between Mo/Si and N atoms.

Table I. Summary of the calculated elastic moduli (E), ideal strengths (σ), critical strains (ε), and failure mechanisms of monolayer $MoSi_2N_4$ and $MoS_2$ under three strain paths

|  | Direction | E (GPa) | σ (GPa) | ε | Failure mechanism |
|---|---|---|---|---|---|
|  | Zigag(x) | 448.3±5.1 | 51.6 | 0.265 | Phonon instability |
| $MoSi_2N_4$ | Armchair(y) | 457.8±3.9 | 49.2 | 0.175 | Elastic instability |
|  | Biaxial | 559.5±7.2 | 50.6 | 0.195 | Phonon instability |
|  | Zigzag | 197.9±4.3 | 15.6 | 0.18 | Elastic instability |
| $MoS_2$[42] | Armchair | 200±3.7 | 24 | 0.256 | Phonon instability |
|  | Biaxial | 250.2±5.8 | 23.8 | 0.195 | Phonon instability |

It is essential to determine whether monolayer $MoSi_2N_4$ remains stable before its maximum stress. To check its stability, we calculate the phonon dispersions, as displayed in Fig. 3. Phonon branches at the stress-free state have no imaginary frequencies showing the dynamical stability of monolayer $MoSi_2N_4$. The three acoustic branches correspond to the vibration in the plane (LA and ZA) and out of the $MoSi_2N_4$ plane (ZA). For the uniaxial tension $\varepsilon_x = 0.265$, a phonon branch has imaginary frequencies along the Γ-Y path as indicated in Fig. 3(b). Examination of the eigenvectors of the unstable phonon modes demonstrates that the imaginary branch is ZA, which corresponds to vibration out of the MoSi$_2$N$_4$ plane. The strength of graphene is also limited by phonon instability. However, the soft phonon branch in such case is in-plane for graphene. Similarly, at the biaxial strain ε=0.195, two phonon branches become unstable (Fig. 3d), and the soft mode at Γ is ZA mode. The cause of failure is the same as those of graphene[39] and monolayer MoS$_2$[42] under biaxial strain, but different from that of silicene. In silicene, the failure is due to the elastic instability[51]. In contrast, all phonon modes remain stable under uniaxial tensile strain along the armchair (y) direction as reaching the maximum tension. That is to say, the failure mechanism of monolayer MoSi$_2$N$_4$ is elastic instability when the tensile strain reaches the ideal strength along the armchair direction. It is worth noting that monolayer MoSi$_2$N$_4$ appears almost isotropic elasticity if we only take into account the elastic moduli and ideal strengths. The different critical strain and phonon dispersions under uniaxial tensions imply that the failure mechanisms upon tensile strains depend on its loading direction and stress state. It is interesting to compare the mechanical response of monolayer MoSi$_2$N$_4$ under tensions with that of MoS$_2$. It found that the failure mechanism of MoSi$_2$N$_4$ under biaxial strain is the same as that of MoS$_2$[42], corresponding to the phonon instability. However, the failure mechanisms of MoSi$_2$N$_4$ under uniaxial tensile strain are different from those of MoS$_2$[42]. For example, the failure mechanisms of MoSi$_2$N$_4$/MoS$_2$ under zigzag and armchair tensile strains are phonon/elastic instability and elastic/phonon instability, respectively. The distinction between monolayer MoSi$_2$N$_4$ and MoS$_2$ in the responses to uniaxial tensions can be attributed to their difference in crystal structures. Monolayer MoS$_2$ is composed of S-Mo-S atomic layers, while monolayer MoSi$_2$N$_4$ exhibits a sandwiched configuration, where a 2H MoS$_2$-type N-Mo-N layer is sandwiched between two SiN layers. The SiN layer has a great influence on the mechanical response of MoSi$_2$N$_4$ monolayer, which will be discussed in detail later.

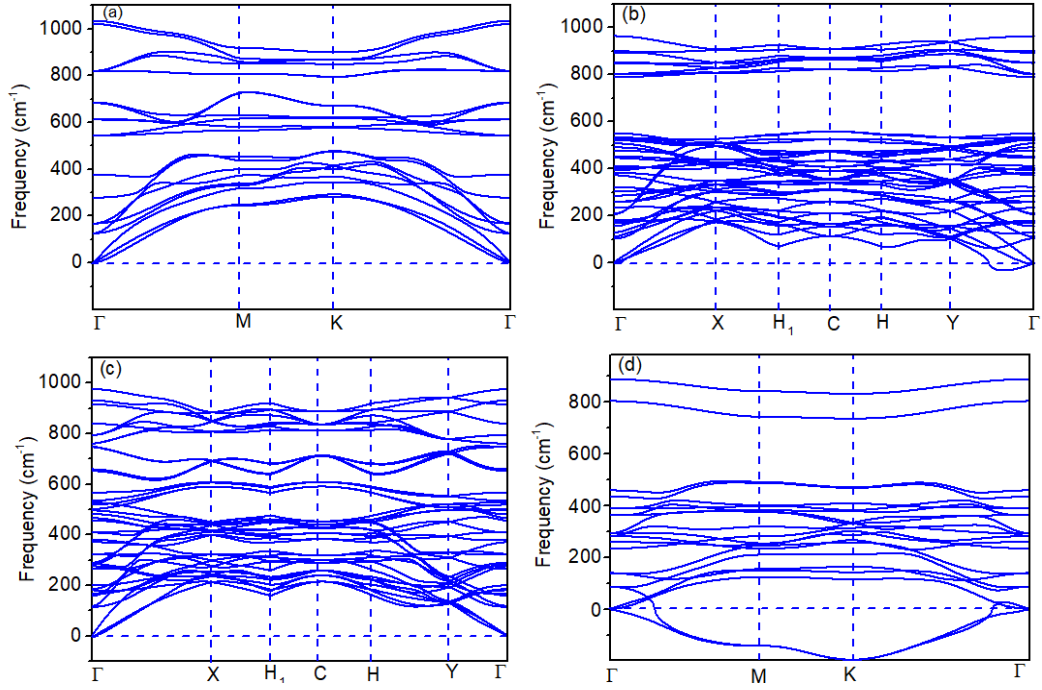

Fig. 3. Phonon dispersions of monolayer MoSi$_2$N$_4$ at (a) stress-free state, (b) uniaxial tensile strain along the x direction ε=0.265, (c) uniaxial tensile strain along the y direction ε=0.175, and (d) biaxial tension with ε=0.195.

The bucking heights are important parameters to characterize the corrugation of 2D buckled materials[44]. We consider three types of bucking heights, which are h$_1$ (height of monolayer MoSi$_2$N$_4$), h$_2$ (height of MoN$_2$), and h$_3$ (height of SiN) shown in Fig.1. It is noted that the tensile strain dependent buckling heights under uniaxial tension are almost isotropic when the stress is less than 0.14. All buckling heights under the biaxial tension and uniaxial tension along the x direction decrease monotonously with increasing tensile strain. Interestingly, all buckling heights under uniaxial y direction tension first decrease and then increase with the increase of strain, especially the buckling height of SiN layer is sensitive to the tensile strain along the y direction. The anomalous change of buckling heights means that the system appears elastically unstable before reaching the maximum tensile strain along the y direction, which is consistent with the above phonon spectrum analysis. The buckling height of MoN$_2$ layer reduces monotonously as increasing tension before reaching the critical strain, therefore the elastic instability under the uniaxial strain along the y direction mainly attributes to the SiN layer.

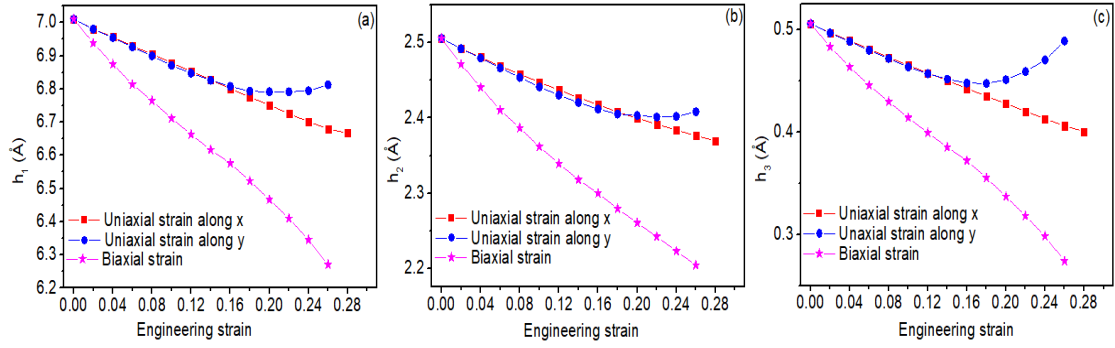

Fig. 4. The dependence of buckling heights under three types of tensile strains. The buckling heights of (a) monolayer MoSi$_2$N$_4$, (b) MoN$_2$ layer, and (c) SiN layer.

To further understand the structural changes of monolayer MoSi$_2$N$_4$ under three types of tensile strains, we calculate the bond lengths of Si-N and Mo-N ($d_1$, $d_2$, $d_3$, $d_4$), and bond angles ($\theta_1, \theta_2, \theta_3, \theta_4$) in Fig. 1 as the functions of the applied strains. When it is applied the tensile strain along the armchair direction, the $d_1$(Si-N) and $d_3$(Mo-N) bonds are perpendicular to the direction of tension. The $d_2$ and $d_4$ bond lengths rise monotonically with the increasing tensile strain, and the $d_1$ and $d_3$ bond lengths slightly reduce, as displayed in Fig. 5(a). The bond angles $\theta_1$ and $\theta_3$ monotonically enhance, while the bond angles $\theta_2$ and $\theta_4$ monotonically decreases with increasing zigzag strain. When the armchair tension is applied (Fig. 5(b)), the $d_1$ and $d_3$ bonds are parallel to the direction of the strain, therefore, their bond lengths elongate with the increase of tension. Surprisingly, the variation of $d_2$ and $d_4$ bond lengths and bond angles with the armchair strain is non-monotonic. The bond length of $d_4$ is stretched slightly at first and then compressed, while the $d_2$ remains more or less constant. The bond angle $\theta_3$ ($\theta_4$) decreases (increases) first, then increases (decreases) when the tensile strain is over 14%, which results from the elastic unstable. The inflection point is consistent with the one of buckling height change. To analyze the evolution of bonded nature near the inflexion point, we also calculate the electron localization function of MoSi$_2$N$_4$ under the free station, the tensile strain of 14%, and critical strain (17.5%) along the y direction. The electron localization is found between Si and N atoms. The $p_z$ states of Si and N are hybridized to form strong σ bonds. The Si$_3$-N bond is along the y direction, thus, the electrons accumulated between Si$_3$ and N atoms become fewer with increasing strain, and the corresponding σ bond becomes weak. When the tension is applied along the y direction, MoSi$_2$N$_4$ should be contracted in the x direction. Then the $Si_1$-N and $S_2$-N bond lengths should become smaller, but the σ bonds between $Si_1$ ($Si_2$) and N are so strong that contraction doesn't happen timely until the stress is greater than 14%. As shown in Fig. 5(c) and (f), both bond lengths and angles under biaxial tension increase and the changes of bond length (bond angle) of $MoN_2$ and SiN layers are similar.

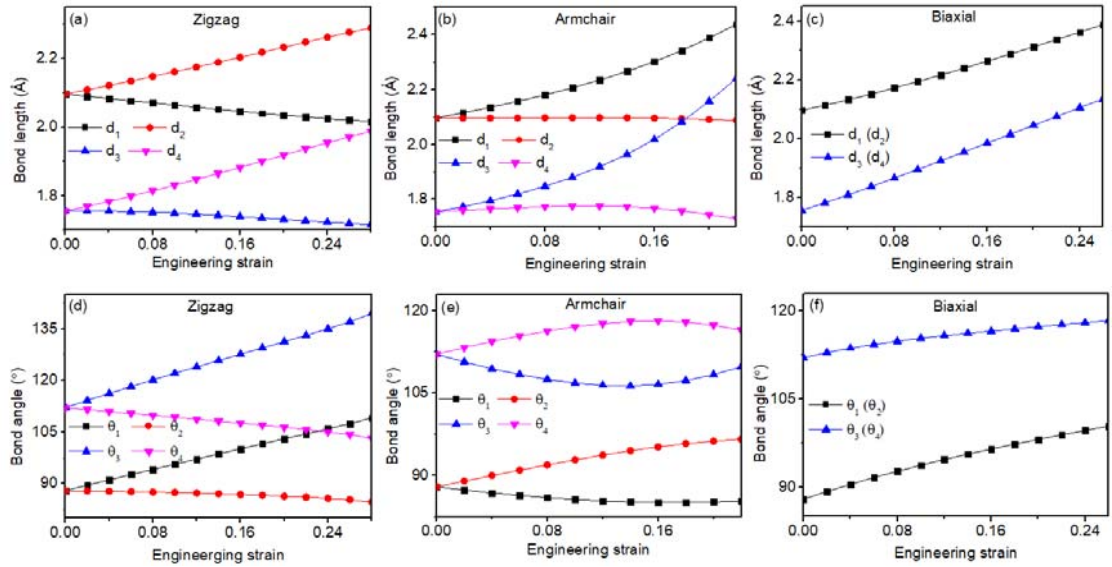

Fig. 5 The dependences of monolayer $MoSi_2N_4$'s bond lengths and bond angles on the (a, d) zigzag, (b, e) armchair uniaxial, and (c, f) biaxial tensions.

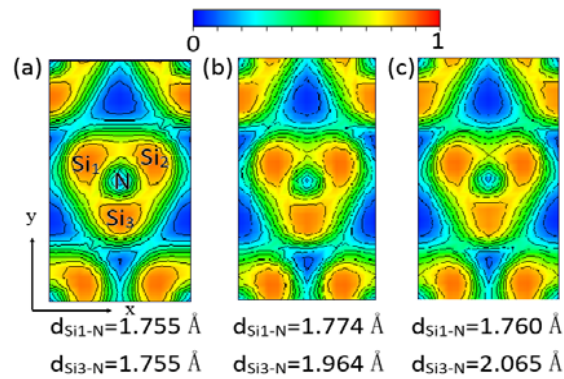

Fig. 6 The electron localization function of Si-N layer in monolayer $MoSi_2N_4$. (a) Unstrained structure, (b) $\varepsilon_y = 0.14$, and (c) $\varepsilon_y = 0.175$.

**CONCLUSIONS**

The mechanical response of monolayer $MoSi_2N_4$ is investigated systematically by using first-principles calculations. The calculated results indicated that monolayer $MoSi_2N_4$ can withstand tensile strength up to 50.6 GPa under biaxial strain, and the corresponding critical strain is 19.5%. The ideal tensile strengths under uniaxial (zigzag and armchair) strains are close to that of biaxial strains. The critical strains of monolayer $MoSi_2N_4$ under zigzag and armchair are 26.5% and 17.5%, which are close to those of monolayer $MoS_2$ under armchair (25.6%) and zigzag (18%) strains. Furthermore, the phonon dispersions are investigated. The results indicate that the failure mechanisms of monolayer $MoSi_2N_4$ are attributed to phonon instability under uniaxial tension along the zigzag direction and biaxial tension. The phonon instability under zigzag tension is attributed to the soft modes of ZA branches. While the failure mechanism along the armchair direction is elastic instability which mainly results from the elastic failure of the SiN layer before reaching the critical strain. The calculated elastic modulus of monolayer $MoSi_2N_4$ is approximately half that of graphene. But its elastic moduli and ideal tensile strengths under uniaxial and biaxial strains are more than twice those of monolayer $MoS_2$. The results mean that monolayer $MoSi_2N_4$ is a hard 2D material. This study can help us to understand the fundamental mechanical properties of monolayer $MoSi_2N_4$, and provide theoretical support for tuning the physical properties by strain engineering.

**ACKNOWLEDGMENTS**

This work is supported by National Natural Science Foundation of China (Grant No. 11704195, 11974163) and the National Key R&D Program of China (Grant No. 2016YFA0201104). QFL also acknowledges the Qing Lan Project of Jiangsu Province (Grant No. R2019Q04). The numerical calculations in this paper have been done on the computing facilities in the High Performance Computing Center (HPCC) of Nanjing University.

[1] K. S. Novoselov, A. K. Geim, S. V. Morozov, D. Jiang, Y. Zhang, S. V. Dubonos, I. V. Grigorieva, A. A. Firsov, Electric field effect in atomically thin carbon films, Science, 306, 666-669 (2004).
[2] Y. Liu, Y. Huang and X. Duan, Van der Waals integration before and beyond two-dimensional materials, Nature, 567(7748), 323-333 (2019).

[3] D. Li, R.B. Kaner, Graphene-based materials, Science, 320, 1170-1171 (2008).

[4] K. Novoselov, A. Mishchenko, A. Carvalho, A. C. Castro Neto, 2D materials and van der Waals heterostructures, Science, 353, aac9439 (2016).

[5] D. Akinwande, C. Huyghebaert, C.H. Wang, M.I. Serna, S. Goossens, L.J. Li, H. S. P. Wong and F. H. L. Koppens, Graphene and two-dimensional materials for silicon technology, Nature, 573(7775), 507-518, (2019).

[6] F. Ersan, D. Kecik, V. O. Özçelik, Y. Kadioglu, O. Üzengi Aktürk, E. Durgun, E. Aktürk, and S. Ciraci, Two-dimensional pnictogens: A review of recent progresses and future research directions, Appl. Phys. Rev, 6(2), 021308, (2019).

[7] B. Anasori, Y. Xie, M. Beidaghi, J. Lu, B. C. Hosler, L. Hultman, P. R. C. Kent, Y. Gogotsi, and M.W. Barsoum. Two-dimensional, ordered, double transition metals carbides (MXenes). ACS Nano, 9, 9507–9516 (2015).

[8] K. Rasool, R. P. Pandey, P. A. Rasheed, S. Buczek, Y. Gogotsi, K.A. Mahmoud, Water treatment and environmental remediation applications of two-dimensional metal carbides (MXenes), Mater. Today, 30, 80–102 (2019).

[9] X. Wang, T. S. Mathis, K. Li, Z. Lin, L. Vlcek, T. Torita, N. C. Osti, C. Hatter, P. Urbankowski, A. Sarycheva, M. Tyagi, E. Mamontov, P. Simon and Y. Gogotsi, Influences from solvents on charge storage in titanium carbide MXenes, Nat. Energy, 4, 241−248 (2019).

[10] A. Sarycheva, A. Polemi, Y. Liu, K. Dandekar, B. Anasori and Y. Gogotsi, 2D titanium carbide (MXenes) for wireless communication, Sci. Adv. 4, eaau0920 (2018).

[11] J. Zhang, Y. Zhao, X. Guo, C. Chen, C.L. Dong, R.S. Liu, C.P. Han, Y. Li, Y. Gogotsi and G. Wang, Single platinum atoms immobilized on an MXene as an efficient catalyst for the hydrogen evolution reaction, Nat. Catal. 1, 985−992 (2018).

[12] X. Xie, N. Zhang, Positioning MXenes in the Photocatalysis Landscape: Competitiveness, Challenges, and Future Perspectives, Adv. Funct. Mater. 30, 2002528 (2020).

[13] B. Anasori, M. R. Lukatskaya, Y. Gogotsi, 2D metal carbides and nitrides (MXenes) for energy storage, Nat. Rev. Mater. 2, 16098 (2017).

[14] J. Pang, R. G. Mendes, A. Bachmatiuk, L. Zhao, H. Q. Ta, T. Gemming, H. Liu, Z. Liu, M.H. Rummeli, Applications of 2D MXenes in energy conversion and storage systems, Chem. Soc. Rev. 48, 72 (2019).

[15] Y. Hong, Z. Liu, L. Wang, T. Zhou, W. Ma, C. Xu, S. Feng, L. Chen, M. Chen, D. Sun, X. Chen, H. Cheng, and W. Ren. Chemical vapor deposition of layered two dimensional $MoSi_2N_4$ materials, Science, 369, 670 (2020).

[16] Y. Cai, G. Zhang, Y.-W. Zhang, Polarity-reversed robust carrier mobility in monolayer $MoS_2$ nanoribbons, J. Am. Chem. Soc. 136, 6269 (2014).

[17] A. Bafekry, M. Faraji, Do M. Hoat, M. M. Fadlallah, M. Shahrokhi, F. Shojaei, D. Gogova, M. Ghergherehchi, $MoSi_2N_4$ single-layer: a novel two-dimensional material with outstanding mechanical, thermal, electronic, optical, and photocatalytic properties, arXiv:2009.04267 (2020).

[18] C. Yang, Z. Song, X. Sun, J. Lu, Valley pseudospin in monolayer $MoSi_2N_4$ and $MoSi_2As_4$, arXiv:2010.10764 (2020).

[19] S. Li, W. Wu, X. Feng, S. Guan, W. Feng, Y. Yao, S. A. Yang, Valley-dependent properties of monolayer $MoSi_2N_4$, $WSi_2N_4$ and $MoSi_2As_4$, Phys.Rev.B, 102, 235435 (2020).

[20] S. D. Guo, Y. T. Zhu, W. Q. Mu and W.C. Ren, Intrinsic piezoelectricity in monolayer $MSi_2N_4$ (M = Mo, W, Cr, Ti, Zr and Hf), EPL, 132(5), 57002 (2020).

[21] X. Guo, S. Guo, Tuning transport coefficients of monolayer $MoSi_2N_4$ with biaxial strain, Chinese Phys. B (2021). In press, https://doi.org/10.1088/1674-1056/abdb22.

[22] L. Cao, G. Zhou, Q. Wang L. K. Ang, Y. S. Ang, Two-dimensional van der Waals electrical contact to monolayer $MoSi_2N_4$, Appl. Phys. Lett. 118, 013106 (2021).

[23] S. D. Guo, Y. T. Zhu, W. Q. Mu, L. Wang and X. Q. Chen, Structure effect on intrinsic piezoelectricity in septuple-atomic-layer $MSi_2N_4$ (M=Mo and W), Comput. Mater. Sci. 188, 110223 (2021).

[24] B. Mortazavi, B. Javvaji, F. Shojaei, T. Rabczuk, A. V. Shapeev, X. Zhuang, Exceptional piezoelectricity, high thermal conductivity and stiffness and promising photocatalysis in two-dimensional $MoSi_2N_4$ family confirmed by first-principles, Nano Energy, 82, 105716 (2020).

[25] J. Yu, J. Zhou, X. Wan, Q. Li, High intrinsic lattice thermal conductivity in monolayer $MoSi_2N_4$, arXiv:2012.14120 (2020).

[26] H. Zhong, W. Xiong, P. Lv, S. Yuan, Strain-induced semiconductor to metal transition in $MA_2Z_4$ bilayers, arXiv. 2009. 09089 (2020).

[27] H. Ai, D. Liu, J. Geng, S. Wang, K. H. Lo, H. Pan, Theoretical evidence of the spin–valley coupling and valley polarization in two-dimensional $MoSi_2X_4$ (X = N, P, and As), Phys. Chem. Chem. Phys. 2021. In press. https://doi.org/10.1039/D0CP05926A.

[28] Q. Yue, S. Chang, J. Kang, S. Qin, and J. Li. Mechanical and electronic properties of graphyne and its family under elastic strain: theoretical predictions, J. Phys. Chem. C, 117(28), 14804-14811 (2013).

[29] C. Lee, X. Wei, J. W. Kysar and J. Hone, Measurement of the elastic properties and intrinsic strength of monolayer graphene, Science, 321, 385–388 (2008).

[30] K. S. Novoselov, V. I. Fal'ko, L. Colombo, P. R. Gellert, M. G. Schwab and K. Kim, A roadmap for graphene, Nature 490, 192–200 (2012).

[31] A. C. Ferrari, F. Bonaccorso, V. Fal'ko, K. S. Novoselov, S. Roche, P. Bøggild, S. Borini, F. H. L. Koppens, V. Palermo, N. Pugno et al, Science and technology roadmap for graphene, related two-dimensional crystals, and hybrid systems, Nanoscale 7, 4598–4810 (2015).

[32] K. Hu, D. D. Kulkarni, I. Choi and V. V. Tsukruk, Graphene-polymer nanocomposites for structural and functional applications, Prog. Polym. Sci. 39, 1934–1972 (2014).

[33] A. Lipatov, H. Lu, M. Alhabeb, B. Anasori, A. Gruverman, Y. Gogotsi, Alexander Sinitskii, Elastic properties

of 2D $Ti_3C_2T_x$ MXene monolayers and bilyaers, Sci. Adv. 4, eaat0491 (2018).

[34] S. Bertolazzi, J. Brivio, A. Kis, Stretching and breaking of ultrathin $MoS_2$, ACS Nano 5, 9703–9709 (2011).

[35] A. Lipatov, M. Alhabeb, H. Lu, S. Zhao, M. J. Loes, N. S. Vorobeva, Y. Dall'Agnese, Y. Gao, A. Gruverman, Y. Gogotsi, A. Sinitskii, Electrical and elastic properties of individual single-layer $Nb_4C_3T_x$ MXene flakes, Adv. Electron. Mater. 6, 1901382 (2020).

[36] T. Li, J.W. Morris, Jr., N. Nagasako, S. Kuramoto, and D. C. Chrzan, "Ideal" engineering alloys, Phys. Rev. Lett. 98 105503 (2007).

[37] J. Pokluda, M. Černý, P. Šandera and M. Šob, Calculations of theoretical strength:State of the art and history, J. Comput.-Aided Mater. Des. 11, 1 (2004).

[38] F. Liu, P. Ming and J. Li, Ab initio calculation of ideal strength and phonon instability of graphene under tension, Phys. Rev. B, 76, 064120 (2007).

[39] C. A. Marianetti and H. G. Yevick, Failure mechanisms of graphene under tension, Phys. Rev. Lett. 105, 245502 (2010).

[40] S. J. Woo and Y. W. Son, Ideal strength of doped graphene, Phys. Rev. B. 87, 075419 (2013).

[41] R.C. Cooper, C. Lee, C. A. Marianetti, X. Wei, J. Hone, and J. W. Kysar, Nonlinear elastic behavior of two-dimensional molybdenum disulfide, Phys. Rev. 87, 035423 (2013)

[42] T. Li, Ideal stength and phonon instability in single-layer MoS2, Phys. Rev. B, 85, 235407 (2012); T. Li, Reply to "Comment on 'Ideal strength and phonon instability in single-layer $MoS_2$' " Phys. Rev. 90, 167402 (2014).

[43] X. Fan, W.T. Zheng, J. L. Kuo and D. J. Singh, Structural stability of single-layer $MoS_2$ under large strain, J. Phys.: Condens. Matter 27, 105401 (2015)

[44] H. Wang, Q. Li, Y. Gao, F. Miao, X. Zhou and X. G. Wan, Strain effects on borophene:ideal strength, negative Possion's ratio and phonon instability, New. J. Phys, 18, 073016 (2016).

[45] Z. Zhang, Y. Yang, E. S. Penev and B.I. Yakobson, Elastivity, flexibility, and ideal strength of borophenes, Adv. Funct. Mater, 27, 1605059 (2017).

[46] Q. Wei and X. Peng, Superior mechanical flexibility of phosphorene and few-layer black phosphorus, Appl. Phys. Lett. 104, 251915 (2014).

[47] J. Tao, W. Shen, S. Wu, L. Liu, Z. Feng, C. Wang, C. Hu, P. Yao, H. Zhang, W. Pang, X. Duan, J. Liu, C. Zhou, and D. Zhang, Mechanical and electrical anisotropy of few-layer black phosphorus, ACS Nano, 9, 11362 (2015).

[48] K. N. Kudin, G. E. Scuseria, and B. I. Yakobson, $C_2F$,BN, and C nanoshell elasticity from ab initio computations, Phys. Rev. B, 64, 235406 (2001).

[49] H. Şahin, S. Cahangirov, M. Topsakal, E. Bekaroglu, E. Akturk, R. T. Senger, and S. Ciraci, Monolayer honeycomb structures of group-IV elements and III-V binary compounds: First-principles calculations, Phys. Rev. B, 80, 155453 (2009).

[50] M. Topsakal and S. Ciraci, Elastic and plastic deformation of graphene, silicene, and boron nitride honeycomb nanoribbons under uniaxial tension: A first-principles density-functional theory study, Phys. Rev. B, 81, 024107 (2010).

[51] C. Yang, Z. Yu, P. Lu, Y. Liu, H. Liu, L. Ye, T. Gao, Phonon instability and ideal strength of silicene under tension, Compu. Mater. Sci, 95, 420-428 (2014).

[52] Z. H. Fu, Q. F. Zhang, D. Legut, C. Si, T. C. Germann, T. Lookman, S. Y. Du, J. S. Francisco, and R. F. Zhang, Stabilization and strengthening effects of functional groups in two-dimensional titanium carbide, Phys. Rev. B, 94, 104103 (2016).

[53] G. Kresse, J. Furthmüller, Efficiency of ab-initio total energy calculations for metals and semiconductors using a plane-wave basis set, Comput. Mater. Sci., 6, 15 (1996).

[54] G. Kresse, J. Furthmüller, Efficient iterative schemes for ab initio total-energy calculations using a plane-wave basis set, Phys. Rev. B, 54, 11169 (1996).

[55] J. P. Perdew, K. Burke and M. Ernzerhof, Generalized gradient approximation made simple, Phys. Rev. Lett., 77, 3865 (1996).

[56] P. E. Blöchl, Projector augmented-wave method, Phys. Rev. B, 50, 17953 (1994).

[57] G. Kresse and D. Joubert, From ultrasoft pseudopotentials to the projector augmented-wave method, Phys. Rev. B, 59, 1758 (1999).

[58] A. Togo, I. Tanaka, First principles phonon calculations in materials science, Scripta. Mater. 108, 1 (2015).